# Optimization of infrared and magnetic shielding of superconducting TiN and Al coplanar microwave resonators

J M Kreikebaum[1], A Dove[1], W Livingston[1], E Kim[1,2], and I Siddiqi[1]

1. Quantum Nanoelectronics Laboratory, Department of Physics, University of California, Berkeley, California 94720, USA

2. Department of Physics, KAIST, Daejoen 34141, Republic of Korea

Email: jmkreikebaum@berkeley.edu

We present a systematic study of the effects of shielding on the internal quality factors ($Q_i$) of Al and TiN microwave resonators designed for use in quantum coherent circuits. Measurements were performed in an adiabatic demagnetization refrigerator, where typical magnetic fields of 200 µT are present at the unshielded sample stage. Radiation shielding consisted of 100 mK and 500 mK Cu cans coated with infrared absorbing epoxy. Magnetic shields consisted of Cryoperm 10 and Sn plating of the Cu cans. A 2.7 K radiation can and coaxial thermalization filters were present in all measurements. TiN samples with $Q_i = 1.3 \times 10^6$ at 100 mK exhibited no significant variation in quality factor when tested with limited shielding. In contrast, Al resonators showed improved $Q_i$ with successive shielding, with the largest gains obtained from the addition of the first radiation and magnetic shields and saturating before the addition of Sn plating infrared absorbing epoxy.

Superconducting quantum circuits are a leading candidate to become the foundation of a fault-tolerant quantum computer, but high coherence materials are required for their successful implementation [1-5]. Dielectric loss, surface participation, excess quasiparticles, and trapped vortices all serve to reduce the internal quality factors ($Q_i$) of superconducting coplanar waveguide (CPW) resonators [6-11] which are frequently used to couple qubits to a transmission line for state readout [12]. In addition to optimizing fabrication recipes to reduce loss, previous work demonstrates the importance of shielding [6]. In this work, we systematically study the effects of shielding TiN and Al CPW resonators in order to develop a robust testing procedure that accurately reflects intrinsic material quality. Moreover, given the space and heat-load constraints of cryogenic instruments, an understanding of optimized shielding is required for the practical implementation of large-scale quantum processors.

The resonators used in this study were fabricated at MIT Lincoln Laboratories. Each 5 x 5 mm die features 5 frequency multiplexed λ/4 resonators capacitively coupled to a single feedline on a Si substrate. Circuits were defined by photolithographically patterning a 5 μm gap around 10 μm wide resonators. TiN resonators were reactively sputtered and dry etched; Al resonators were evaporated and wet etched. The quality factor of the coupling between the resonator and the transmission line ($Q_e$) was designed to be less than the $Q_i$ of the resonators ($Q_e \approx 3 \times 10^5$) and perforations were patterned into the ground plane to serve as flux pinning sites. Further fabrication details will be published by MIT at a later time.

The experimental setup is depicted in figure 1. Samples were mounted to the coldest stage of an adiabatic demagnetization refrigerator and were surrounded by successive layers of magnetic and infrared shielding whose effects were studied systematically. All measurements were performed at 100 mK. The stainless steel microwave input line was thermally anchored and attenuated at all temperature stages terminating with a 50 Ω lossy filter with 3.5 dB of attenuation at 5 GHz to ensure thermalization of the coaxial lines and to absorb infrared photons[13]. At the coldest stage, Sn plated Cu coaxial cable was used and the total attenuation of the input line was 70 dB at room temperature. After the sample, two circulators were used to provide approximately 36 dB of isolation from microwaves propagating towards the sample from the output line. The signal then passed through another lossy filter before entering a Nb coax leading to a LNF-LNC4_8A HEMT at 2.7 K. A vector network analyzer (VNA) recorded transmission spectra ($S_{21}$) after passing through an additional room temperature amplifier. The resulting trace was fit using a least squares optimization described in [14].

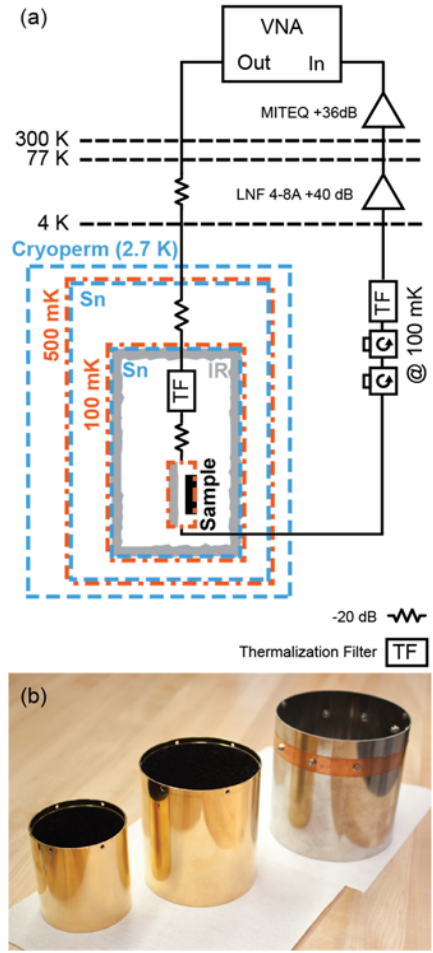

Figure 1. (Color online) (a) Fully shielded experimental set up: magnetic shields are shown in blue-dashed lines, thermal shields are shown in red-dash-dot lines. The sample is mounted in a Cu box with infrared absorbing epoxy (IR) opposite the sample. Cu cans at 100 mK and 500 mK serve as infrared radiation shielding. The 100 mK can also has infrared absorbing epoxy (shown in grey). Sn plating on the Cu cans and cryoperm at 2.7 K provides magnetic shielding. Lossy thermalization filters (TF) are present on the input and output lines. (b) Picture of the shields used. From left to right: 100 mK Cu can, 500 mK Cu can, and cryoperm.

Experimental data from the shielding study for Al CPW resonators are shown in figure 2. The sample was first mounted in a minimally shielded configuration consisting of a wirebonded chip in a Cu cryopackage with no lid such that the sample had no magnetic shielding and was directly exposed to 2.7

K infrared radiation. With this configuration, we measured a mean internal quality factor $\overline{Q_i} = 1.71 \times 10^5$ at single photon powers ($\overline{n} \approx 1$). The addition of an initial layer of infrared shielding in the form of a cryopackage lid approximately doubled $\overline{Q_i}$. Removing the lid and adding cryoperm at 2.7 K as an initial layer of magnetic shielding also resulted in $\overline{Q_i}$ doubling. The cryoperm is nominally expected to reduce the magnetic field by a factor of ~1500. Since these two shields are protecting the Al CPW resonator from different loss mechanisms, we expect that their effects would combine linearly, and indeed the data show an improvement in $\overline{Q_i}$ by a factor of 3.5 over the unshielded configuration. These systematic improvements confirm the trends reported in the literature [6], and we find that Al CPW resonators benefit from both infrared shielding and reduction in ambient magnetic fields.

With a closed cryopackage and cryoperm, we obtain $\overline{Q_i} = 5.9 \times 10^5$. By adding subsequent layers of infrared radiation shielding in the form of two Cu cans, at 100 mK and 500 mK, which were designed to be nearly-light tight, we see a small increase to $\overline{Q_i} = 6.45 \times 10^5$. Repeated measurement in this configuration three months later exhibited no degradation, but rather a slight increase in $\overline{Q_i}$ representative of run-to-run scatter. Adding shielding beyond this configuration gives a null result where the variation of the sample with time is larger than any improvement due to the shielding. This small drift in $Q_i$ is not yet understood, and can also be seen by repeating many measurements within a single cooldown.

The shields that were shown to have little effect on $\overline{Q_i}$ for Al CPW resonators were a 2 μm layer of Sn, a superconducting magnetic shield, and an ~2 mm thick coating of infrared absorbing epoxy on the inside of the 100 mK Cu can and on the surface of the cryopackage lid facing the sample. The Sn was electroplated onto the Cu radiation cans before Au plating. Devices tested with the Sn plating but without the cryoperm performed the same as having no magnetic shielding, suggesting that this specific type of shield is not effective. The infrared absorbing epoxy consisted of, by mass, 68% Stycast 2850 LT, 5% Catalyst 24LV, 7% Carbon lampblack, and 20% 350 μm SiC grit [15]. The lack of improvement in $\overline{Q_i}$ after adding this epoxy suggests there is no infrared light within the 100 mK radiation can. To test our hypothesis, we removed the cryopackage lid but kept all other shields, resulting is a slightly decreased $\overline{Q_i}$, but broader range of $Q_i$, and the highest $Q_i$ measured for Al CPW resonators in the shielding study at single photon powers, supporting the claim that resonators are not infrared light limited. The dominant source of drift in $\overline{Q_i}$ appears to be cooldown to cooldown variations and infrared absorbing epoxy and Sn plating do not result in improvement greater than this drift.

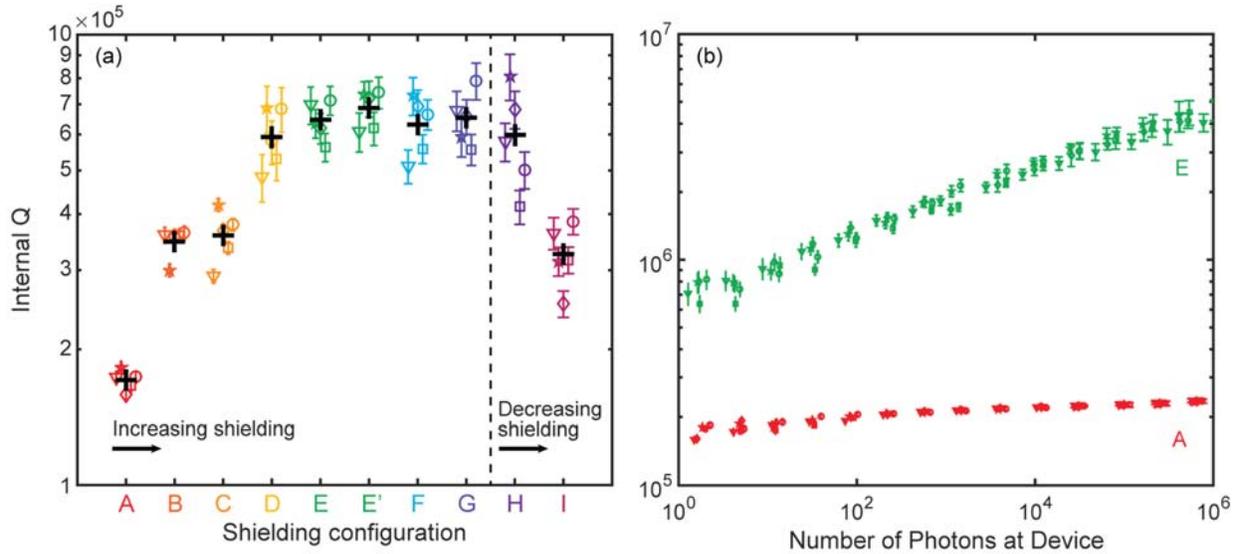

Figure 2. (color online) (a) Single photon power $Q_i$ of Al CPW resonators at 100 mK in various shielding configurations for the five resonators on the die. The error bars represent the uncertainty of the least squares fit and the black + is the average of the five resonators. A: Directly exposed to 2.7 K radiation in a 200 µT ambient field. B: With cryopackage lid. C: No lid, but with cryoperm at 2.7 K. D: Lid and cryoperm. E & E': Lid, cryoperm, and 500 & 100 mK Cu cans repeated three months apart. F: With Sn on Cu cans. G: With infrared absorbing epoxy on lid and Cu cans. H: Kept all shields, except cryopackage lid. I: Kept all shields, except cryoperm. (b) $Q_i$ of all 5 resonators as a function of power in configurations A and E.

For TiN, a strikingly different result emerges from the same systematic shielding study, seen in figure 3. Again, starting with the sample directly exposed to 2.7 K radiation, and then adding successive shields, we see that $\overline{Q_i}$ does not improve in a systematic way, but rather variations appear to be attributable to measurements being performed at different times, between which the sample is heated and cooled. Stray infrared light from higher temperature stages can cause excess quasiparticle generation in superconducting films, and Al is particularly susceptible to this effect since the superconducting gap is small and quasiparticle recombination times are slow [6, 16]. TiN has a 4.6 times larger superconducting gap, resulting in a predicted 46% reduction in quasiparticles generated from the same background radiation. In addition, TiN's lack of response to reductions in ambient magnetic fields by factors of ~1500 suggests that these 10 μm wide TiN resonators have a threshold higher than 200 μT to expel all trapped magnetic flux [7, 17].

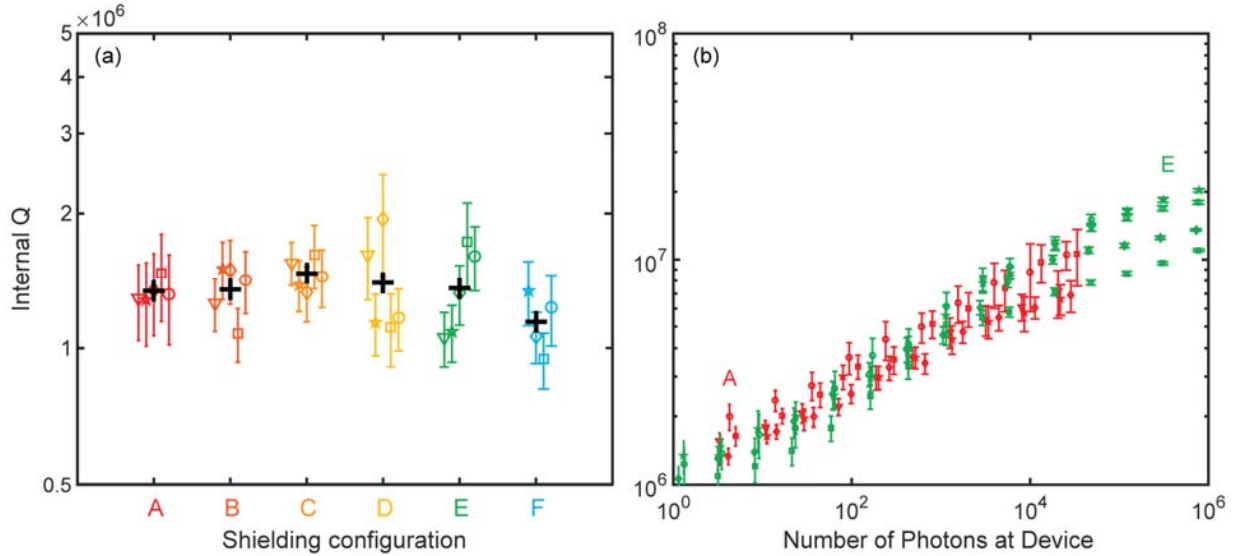

Figure 3. (color online) (a) Single photon power $Q_i$ of TiN CPW resonators at 100 mK in various shielding configurations for the five resonators on the die. The error bars represent the uncertainty of the least squares fit and the black + is the average of the five resonators. A: Directly exposed to 2.7 K radiation in a 200 μT ambient field. B: With lid. C: With 100 & 500 mK Cu cans D: With cryoperm. E: Sn plate Cu cans. F: With infrared absorbing epoxy on Cu cans and surface of lid opposite the sample. A lack of systematic gain suggests that the film is strongly decoupled from loss due to flux vortices and quasiparticle generation in the films. (b) $Q_i$ of all 5 resonators as a function of power in configurations A and E.

In conclusion, we have presented a systematic study of the effect on $\overline{Q_i}$ with various levels of infrared and magnetic shielding around Al and TiN CPW resonators in an adiabatic demagnetization refrigerator. We have found that magnetic and infrared radiation shielding is key to optimal performance for Al CPW devices. A superconducting shield should further reduce ambient magnetic fields around the sample compared to cryoperm alone, but 2 μm of electroplated Sn is insufficient. The effect of adding infrared absorbing epoxy is smaller than the variation in $\overline{Q_i}$ cooldown to cooldown and our inner radiation can is just as effective as a cryopackage lid for blocking infrared radiation. TiN, a material with a significantly larger superconducting gap, shows no dependence of $Q_i$ on added shielding. Devices exhibit nearly identical behavior when exposed directly to 2.7 K radiation and 200 μT magnetic fields as when they are maximally shielded. Further experiments will be performed to see if this same behavior occurs with transmon qubits made of Al coupled to TiN resonators.

We thank our collaborators at MIT Lincoln Labs for supplying samples for this study. This work was supported by the Army Research Office under Grant # W911NF-15-1-0496.

1. Devoret, M.H. and J.M. Martinis, *Implementing Qubits with Superconducting Integrated Circuits.* Quantum Information Processing, 2004. **3**(1): p. 163-203.
2. Clarke, J. and F.K. Wilhelm, *Superconducting quantum bits.* Nature, 2008. **453**(7198): p. 1031-42.
3. Schoelkopf, R.J. and S.M. Girvin, *Wiring up quantum systems.* Nature, 2008. **451**(7179): p. 664-669.
4. Makhlin, Y., G. Schön, and A. Shnirman, *Quantum-state engineering with Josephson-junction devices.* Reviews of Modern Physics, 2001. **73**(2): p. 357-400.
5. DiVincenzo, D.P., *Fault-tolerant architectures for superconducting qubits.* Physica Scripta, 2009. **T137**: p. 014020.
6. Barends, R., et al., *Minimizing quasiparticle generation from stray infrared light in superconducting quantum circuits.* Applied Physics Letters, 2011. **99**(11): p. 113507.
7. Song, C., et al., *Microwave response of vortices in superconducting thin films of Re and Al.* Physical Review B, 2009. **79**(17).
8. Wenner, J., et al., *Surface loss simulations of superconducting coplanar waveguide resonators.* Applied Physics Letters, 2011. **99**(11): p. 113513.
9. Wang, H., et al., *Improving the coherence time of superconducting coplanar resonators.* Applied Physics Letters, 2009. **95**(23): p. 233508.
10. Wang, C., et al., *Surface participation and dielectric loss in superconducting qubits.* Applied Physics Letters, 2015. **107**(16): p. 162601.
11. Quintana, C.M., et al., *Characterization and reduction of microfabrication-induced decoherence in superconducting quantum circuits.* Applied Physics Letters, 2014. **105**(6): p. 062601.

12. Wallraff, A., et al., *Strong coupling of a single photon to a superconducting qubit using circuit quantum electrodynamics.* Nature, 2004. **431**(7005): p. 162-167.
13. Birenbaum, J., *The C-shunt Flux Qubit: A New Generation of Superconducting Flux Qubit.* 2014.
14. Khalil, M.S., et al., *An analysis method for asymmetric resonator transmission applied to superconducting devices.* Journal of Applied Physics, 2012. **111**(5): p. 054510.
15. Persky, M.J., *Review of black surfaces for space-borne infrared systems.* Review of Scientific Instruments, 1999. **70**(5): p. 2193-2217.
16. de Visser, P.J., et al., *Number fluctuations of sparse quasiparticles in a superconductor.* Phys Rev Lett, 2011. **106**(16): p. 167004.
17. Nsanzineza, I. and B.L. Plourde, *Trapping a single vortex and reducing quasiparticles in a superconducting resonator.* Phys Rev Lett, 2014. **113**(11): p. 117002.